\documentclass[twocolumn,aps,pra,showpacs,amsmath,superscriptaddress]{revtex4}
\usepackage{amssymb}
\usepackage{mathrsfs}
\usepackage{graphicx}
\usepackage{float}

\begin{document}

\title{Superfluid to Bose-glass transition of hard core bosons in
one-dimensional incommensurate optical lattice}

\author{Xiaoming Cai}
\affiliation{Beijing National Laboratory for Condensed Matter
Physics, Institute of Physics, Chinese Academy of Sciences,
Beijing 100190, China}
\author{Shu Chen}
\affiliation{Beijing National
Laboratory for Condensed Matter Physics, Institute of Physics,
Chinese Academy of Sciences, Beijing 100190, China}
\author{Yupeng Wang}
\affiliation{Beijing National Laboratory for Condensed Matter
Physics, Institute of Physics, Chinese Academy of Sciences, Beijing
100190, China}
\date{ \today}

\begin{abstract}
We study superfluid to Anderson insulator transition of strongly
repulsive Bose gas in a one dimensional incommensurate optical
lattice. In the hard core limit, the Bose-Fermi mapping allows us
to deal with the system exactly by using the exact numerical
method. Based on the Aubry-Andr\'{e} model, we exploit the phase
transition of the hard core boson system from superfluid phase
with all the single particle states being extended to the Bose
glass phase with all the single particle states being Anderson
localized as the strength of the incommensurate potential
increasing relative to the amplitude of hopping. We evaluate the
superfluid fraction, the one particle density matrices, momentum
distributions, the natural orbitals and their occupations. All of
these quantities show that there exists a phase transition from
superfluid to insulator in the system.
\end{abstract}


\maketitle

\section{Introduction}

Anderson localization \cite{Anderson} was predicted fifty years
ago as the localization of the electronic wave function in a
disordered potential. So far the phenomenon of Anderson
localization has been observed in variety of systems, such as
electromagnetic waves \cite{Wiersma,Dalichaouch}, sound waves
\cite{Weaver}, and quantum matter waves
\cite{Billy,Roati,Chabe,Edwards}. Recently lots of attentions have
been payed on the cold atom systems in disordered potentials
\cite{Modugno,Adhikari} because of the experimental realization of
the Anderson localization of the quantum matter waves
\cite{Billy,Roati}. The good tunability in ultracold atom systems
\cite{Bloch} offers myriad opportunities for studying the disorder
effects in a controllable way. Random potential can be introduced
by baser beams generating speckle patterns in the optical lattice
\cite{Billy,Lye}. By this way, Billy {\it et al.} observed
exponential localization of a Bose-Einstein condensate (BEC) of
$^{87}$Rb atoms released into a one-dimensional waveguide
\cite{Billy}. Also one can achieve the random potential by loading
a mixture of two kinds of atoms in a optical lattice with one
heavy and one light, and the effect of the heavy ones is to
produce an effective random potential for the lighter atoms when
the heavy ones localize in the lattice randomly \cite{Gavish}.
Another way to produce the random potential is to superimpose two
one-dimensional (1D) optical lattices with incommensurate wave
lengths to generate quasi-periodic potential \cite{Roati,Fallani}.
Particularly, Roati \emph{et al}. \cite{Roati} observed
localization  of a noninteracting BEC of $^{39}$K atoms in a 1D
system with two optical lattice potentials.

On the other hand, the interactions between bosons in the cold atom
systems can be controllably tuned by Feshbach resonances. It is thus
feasible to experimentally study \cite{Roati1} the interplay between
disorder and interactions in bosonic systems under controlled
conditions, which has been a longstanding problem subjected to
intensive studies. Theoretically, it has been predicted that there
is a phase transition from a superfluid phase with extended single
particle states to the Bose-glass (BG) phase
\cite{Giamarchi,Fisher1,Delande,Fontanesi} with Anderson localized
single particle states. The study of the interplay of disorder and
interaction in strongly interacting ultracold atomic systems also
gets lots of attentions \cite{Orso,Roux,Deng,Roscilde,Gurarie}.
Despite the intensive studies
\cite{Orso,Roux,Deng,Roscilde,Gurarie,Scalettar,Zhang,Singh,Rapsch},
the properties of the disordered interacting bosons remain on
debate. The theory of disordered interacting bosons is difficult and
exact solutions are rarely known apart from numerical or approximate
results. An exception is in the one-dimensional (1D) limit with
infinitely strong repulsive interactions, where the disordered boson
problem can be exactly solvable via a Bose-Fermi mapping
\cite{Egger}. In the present work, we study the interacting bosons
in the incommensurate optical lattices in the limit with infinitely
repulsive interaction. In this limit, the Bose gas is known as
Tonks-Girardeau gas or hard core bosonic gas \cite{Girardeau}, which
has been experimentally realized \cite{Toshiya,Paredes}. The
Bose-Fermi mapping or the Jordan-Winger transformation for hard-core
bosons in 1D optical lattice establishes a connection to
non-interacting disordered fermions, which greatly simplify the
computation of physical quantities like the momentum distribution.
Following an exact numerical approach proposed by Rigol and
Muramatsu \cite{Rigol}, we calculate the superfluid fractions, one
particle density matrices, momentum distributions, natural orbitals
and their occupations to exploit the phase transition of the system
from superfluid to BG phase.

\section{Model}

We consider $N$ hard-core ultra-cold bosons in a 1D incommensurate
optical lattice. Under the single band tight binding approximation,
the system can be described by the following hard core bosons
Hamiltonian:
\begin{equation}
\label{eqn2}
H=-t\sum_i(b^\dagger_ib_{i+1}+H.c.)+V\sum_i\mathrm{cos}(\alpha2\pi
i+\delta)n^b_i
\end{equation}
where $b^\dagger_i(b_i)$ is the creation (annihilation) operator of
the boson and they satisfy the hard core constraints \cite{Rigol},
{\it i.e.,}  the on-site anti-commutation $(\{b_i,b^\dagger_i\}=1)$
and $[b_i,b^\dagger_j]=0$ for $i\neq j$; $n^b_i$ is the bosonic
particle number operator; $t$ is the amplitude of hopping and we
will set it to be the unit of the energy $(t=1)$; $V$ is the
strength of incommensurate potential with $\alpha$ being an
irrational number characterizing the degree of the
incommensurability and $\delta$ an arbitrary phase (in our
calculation it is chosen to be zero for convenience, without loss of
generality).

In order to obtain the ground state properties of the hard core
bosons, we use the Jordan-Wigner transformation \cite{Jordan}
\begin{equation}
b^\dagger_j=f^\dagger_j\prod^{j-1}_{\beta=1}e^{-i\pi f^\dagger_\beta
f_\beta},b_j=\prod^{j-1}_{\beta=1}e^{+i\pi f^\dagger_\beta
f_\beta}f_j,
\end{equation}
which maps the hard core bosons Hamiltonian into a noninteracting
spinless fermions Hamiltonian
\begin{equation}
\label{eqn1}
H=-\sum_i(f^\dagger_if_{i+1}+H.c.)+V\sum_i\mathrm{cos}(\alpha2\pi
i)n^f_i
\end{equation}
where $f^\dagger_i(f_i)$ is the creation (annihilation) operator of
the spinless fermion and $n^f_i$ is the particle number operator.
The ground state wave function of the $N$ spinless free fermionic
system can be obtained by diagonalizing Eq.(\ref{eqn1}) and be
written as:
\begin{equation}
\label{eqn3}
|\Psi^G_F\rangle=\prod^N_{n=1}\sum^L_{i=1}P_{in}f^\dagger_i|0\rangle
\end{equation}
where $L$ is the number of the lattice sites, $N$ is the number of
fermions (same as bosons), and the matrix $P$ is given by the
lowest $N$ eigenfunctions of the Hamiltonian,
\begin{equation}
P=\left(
\begin{array}{cccc}
P_{11} & P_{12} & \cdots & P_{1N} \\
P_{21} & P_{22} & \cdots & P_{2N} \\
\vdots & \vdots &  & \vdots \\
P_{L1} & P_{L2} & \cdots & P_{LN}
\end{array}
\right)
\end{equation}
with all the rows being ordered starting from the one with the
lowest energy.

The solution of $P_{in}$ reduces to a single particle problem. The
motion of a single particle in a 1D lattice can be described by the
nearest neighbor tight binding equation
\begin{equation}
\label{eqn4} u_{i+1,n}+u_{i-1,n}+V_i u_{i,n}=E_n u_{i,n},
\end{equation}
where the amplitude of the nearest neighbor hopping has been set to
be unity, $u_{i,n}=P_{in}$ is the amplitude of the particle wave
function at $i$ site with $V_i$ the on site diagonal potential and
$E_n$ is the $n$-th single particle eigen energy with the eigen
state given by $|n\rangle=\sum_{i} u_{i,n} f_{i}^{\dagger}
|0\rangle$. Obviously the solution to Eq. (\ref{eqn4}) is dependent
on the potential $V_i$. For periodic $V_i$ Bloch's theorem tells us
that all the single particle states are extended band states. For a
random $V_i$ the system is the Anderson model and Eq.(\ref{eqn4})
produces \cite{Thouless} localized states in 1D. For the
incommensurate case $V_i=V\mathrm{cos}(2\pi\alpha i)$ with
irrational $\alpha$, Eq.(\ref{eqn4}) is the well-known
Aubry-Andr\'{e} model \cite{Aubry}. Aubry and Andr\'{e} showed that
when $V<2$ all the single particle states are extended and when
$V>2$ all the single particle states are the localized states.
Mobility edges also have been found for the extended Aubry-Andr\'{e}
model \cite{Sarma,Biddle}.

\section{Transition from superfluid to Bose-glass phase}
\begin{figure}[tbp]
\includegraphics[scale=0.8]{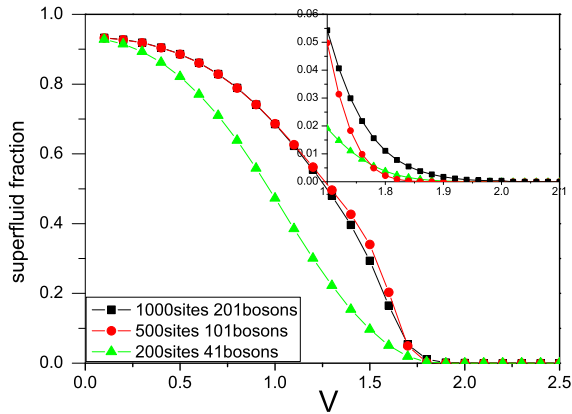}
\caption{(Color online)The superfluid fraction versus $V$ (the
strength of the incommensurate potential) for three different
lattice sizes and fillings with $\alpha=(\sqrt{5}-1)/2$. Insert:
enlargement of picture around the critical point.} \label{Fig1}
\end{figure}

To see whether a bosonic system is in a superfluid phase, we can
calculate the superfluid fraction, which reflects the response of a
superfluid to the imposed phase gradient. Generally a nonzero
superfluid fraction $f^N_s$ characterizes the system being in the
superfluid phase. The superfluid fraction can be calculated with the
aid of Peierls phase factors introduced in the Hamiltonian
Eq.(\ref{eqn2}) by means of the replacement
$b^\dagger_ib_{i+1}\rightarrow b^\dagger_ib_{i+1}e^{i\varphi}$. In
terms of free energy $F_N(\varphi)$ for very small $\varphi$, the
superfluid fraction is then determined as \cite{Roth,Fisher}
\begin{equation}
f^N_s=\underset{\varphi\rightarrow0}{\mathrm{lim}}\tfrac{F_N(\varphi)-F_N(0)}{t\varphi^2N}.
\end{equation}
For hard core bosons in one dimension, the above expression can be
represented as \cite{Krutitsky}
\begin{eqnarray}
f^N_s&=&\tfrac{1}{2N}\sum^L_{i=1}\sum^N_{n=1}[u^*_{i,n} u_{i+1,n}+c.c ]\notag\\
&&-\tfrac{1}{N}\sum^L_{n=N+1}\sum^N_{m=1}\tfrac{1}{E_n-E_m}|\sum^L_{i=1}[u^*_{i,n}
u_{i+1,m} \notag\\
&&-u^*_{i+1,}u_{i,m}]|^2,
\end{eqnarray}
with $u_{L+1,n}=(-1)^{N+1}u_{1,n}$ and $E_n$ is the eigenenergy of
the $n$-th row in $P$ matrix. The results of numerical calculations
for the superfluid fractions for the hard core bosons in
incommensurate lattices are shown in Fig.\ref{Fig1} for three
different lattice sizes and fillings. Here we consider odd numbers
of bosons and use periodic boundary conditions \cite{Lieb}.  When
the strength of the incommensurate potential $V$ is small, the
system is in the superfluid phase with nonzero superfluid fraction.
As $V$ increases, the incommensurate potential acting as a
pseudo-random potential makes the bosons difficult to hop and tend
to be localized. Consequently, the superfluid fraction decreases.
When $V$ gets around $2$, the superfluid fraction approaches zero,
and the system is in the Bose glass phase (insulating phase) with
all the bosons being localized \cite{Fontanesi}.
We note that with different irrational $\alpha$ and fillings the
curves of superfluid fraction exhibit similar behavior.

\begin{figure}[tbp]
\includegraphics[scale=0.8]{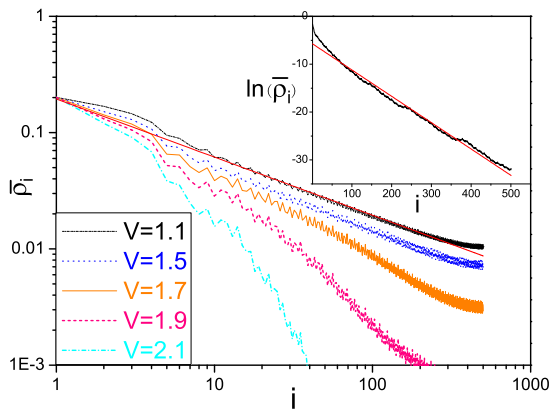}
\caption{(Color online) Log-Log plots of the mean one particle
density matrices for different $V$ of the systems with 201 bosons
on 1000 lattice sites and $\alpha=(\sqrt{5}-1)/2$. The red
straight line shows a power-law fit
($\overline{\rho}_i=0.195/\sqrt{i}$) for the mean one particle
density matrix with $V=1.1$. Insert: The log-line plot of the mean
one particle density matrix for $V=2.1$, and the red straight line
shows an exponential-law fit
($\mathrm{ln}\overline{\rho}_i=-5.697-0.055i$) for the mean one
particle density matrix.} \label{Fig2}
\end{figure}

The one particle Green's function for the hard core bosons can be
written in the form
\begin{eqnarray}
G_{ij}&=&\langle\Psi^G_{HCB}|b_ib^\dagger_j|\Psi^G_{HCB}\rangle\\
&=&
\langle\Psi^A|\Psi^B\rangle\notag
\end{eqnarray}
where $|\Psi^G_{HCB}\rangle$ is the ground state of hard core
bosons, and $\langle\Psi^A| =
\left(f^\dagger_i\prod_{\beta=1}^{i-1}e^{-i\pi f^\dagger_\beta
f_\beta}|\Psi^G_F\rangle\right)^\dagger$, $ |\Psi^B\rangle
=f^\dagger_j\prod_{\gamma=1}^{j-1}e^{-i\pi f^\dagger_\gamma
f_\gamma}|\Psi^G_F\rangle $. After a straightforward evaluation, the
state $\left| \Psi ^A\right\rangle $ can be written as
\[
\left| \Psi^A\right\rangle
=\prod_{n=1}^{N+1}\sum_{l=1}^LP_{ln}^{A}f_l^{\dagger }\left|
0\right\rangle
\]
with $ P_{ln}^{A} = - P_{ln} $ for $l\leq i-1$,
$P_{ln}^{A} =P_{ln}$ for $l\geq i$ with $n\leq N$, $P_{iN+1}^{A}=1$ and $P_{lN+1}^{A}=0$ $%
\left( l\neq i\right) $, i.e., $P^{A}$ is formed by changing the
sign of the elements $P_{ln}$ for $l<i$ because of the exponential
term of the Jordan-Wigner transformation, and adding a column to $P$
with element $P_{iN+1}=1$ and all the others equal to zero for the
further creation of a particle at site $i$. In the same way we can
get $P^{B}$ for the state $|\Psi^B\rangle$ which has the same form
as $\left| \Psi ^A\right\rangle $ with the replacement of $i$ by
$j$. The Green's function is a determinant dependent on the $L\times
\left( N+1\right) $ matrices $P^{A}$ and $P^{B}$ \cite{Rigol}
\begin{eqnarray}
G_{ij}=\left\langle \Psi^A|\Psi^B\right\rangle =\det \left[ \left(
P^{A}\right)^\dagger P^{B}\right] .
\end{eqnarray}
It follows that the one particle density matrix can be determined by
the expression
\begin{equation}
\rho_{ij}=\langle
b^\dagger_ib_j\rangle=G_{ij}+\delta_{ij}(1-2G_{ii}).
\end{equation}
Then we define the mean one particle density matrix as
\begin{equation}
\overline{\rho}_{i}=\tfrac{1}{L}\sum^L_{j=1}\rho_{j,j+i-1}
\end{equation}
to reduce the drastic oscillations caused by the incommensurate
potential. The mean one particle density matrices for different $V$
are shown in Fig.\ref{Fig2}. The mean one particle density matrices
for $V<1.1$ are almost the same as the one with $V=1.1$, having the
power-law decay with exponent of $-1/2$, which has the same exponent
of the hard core bosons in the lattice without the incommensurate
potential \cite{Rigol}. As $V$ increases further, the mean one
particle density matrix still has power-law decay, but the exponent
is smaller than $-1/2$. Correspondingly the superfluid fraction
decreases fast (see Fig.\ref{Fig1}). When $V$ exceeds the critical
point, the mean one particle density matrix has the exponential-law
decay which is the character of the Bose glass phase.

\begin{figure}[tbp]
\includegraphics[width=8.4cm, height=7.5cm, bb=25 20 303 235]{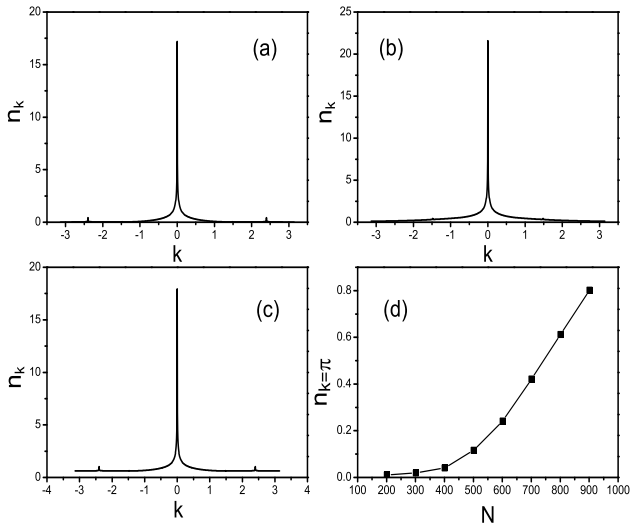}
\caption{Momentum distributions for systems with 1000 lattice
sites, $\alpha=(\sqrt{5}-1)/2$, V=1.1, and fillings of 201 (a),
501 (b), 801 (c) hard core bosons. (d): The value of the tail of
momentum distributions relates to the number of bosons ($N$) for
systems with 1000 lattice sites, $\alpha=(\sqrt{5}-1)/2$ and
V=1.1.} \label{Fig3}
\end{figure}

The momentum distribution is defined by the Fourier transform with
respect to $i-j$ of the one particle density matrix with the form
\begin{equation}
n(k)=\tfrac{1}{L}\sum^L_{i,j=1}e^{-ik(i-j)}\rho_{ij} ,
\end{equation}
where $k$ denotes the momentum. In Fig. \ref{Fig3} we show the
momentum distributions for systems with three different fillings in
1000 lattice sites. Odd numbers of particles are taken for periodic
boundary conditions. The peak structure in the momentum
distributions reflects the bosonic nature of the particles, which is
in contrast with the structure of the momentum distributions of the
equivalent noninteracting fermions. Also because of the existence of
the additional lattice to produce the incommensurate potential, it
is possible that there exist other peaks besides at $k=0$ in the
momentum distributions. For $\alpha=(\sqrt{5}-1)/2$, the secondary
peaks are found to appear at $k=\pm(\alpha-1)2\pi$ only for low and
high fillings. On increasing the number of the bosons the population
of the high momenta states is always increasing, which leads to the
increase in the tails of the momentum distribution as shown in
Fig.\ref{Fig3}d.
\begin{figure}[tbp]
\includegraphics[width=8.4cm, height=7.5cm, bb=25 20 303 235]{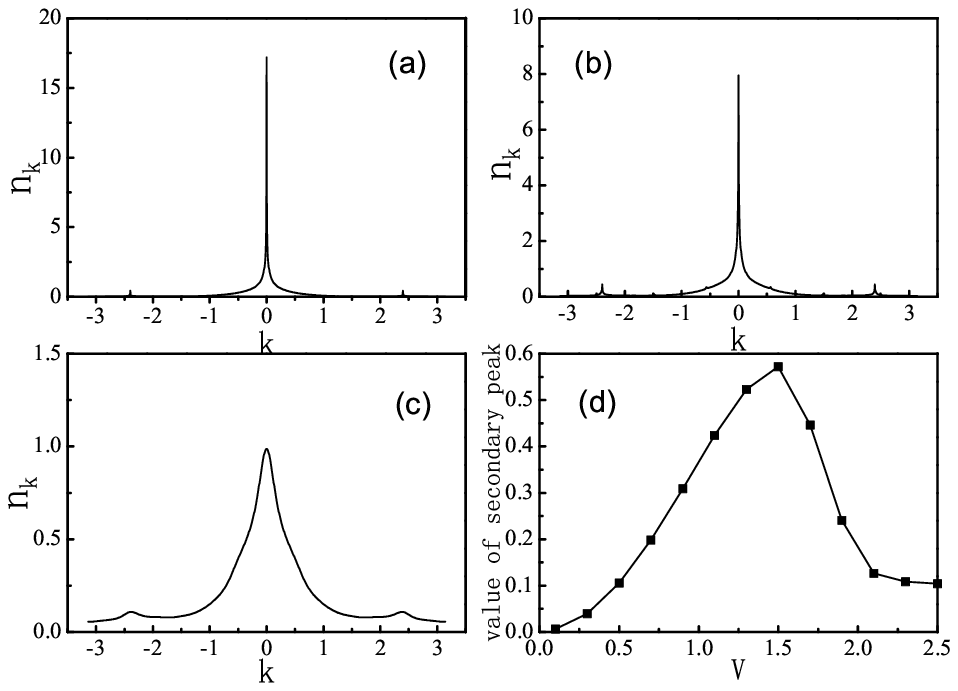}
\includegraphics[width=5cm, height=4cm, bb=25 20 303 235]{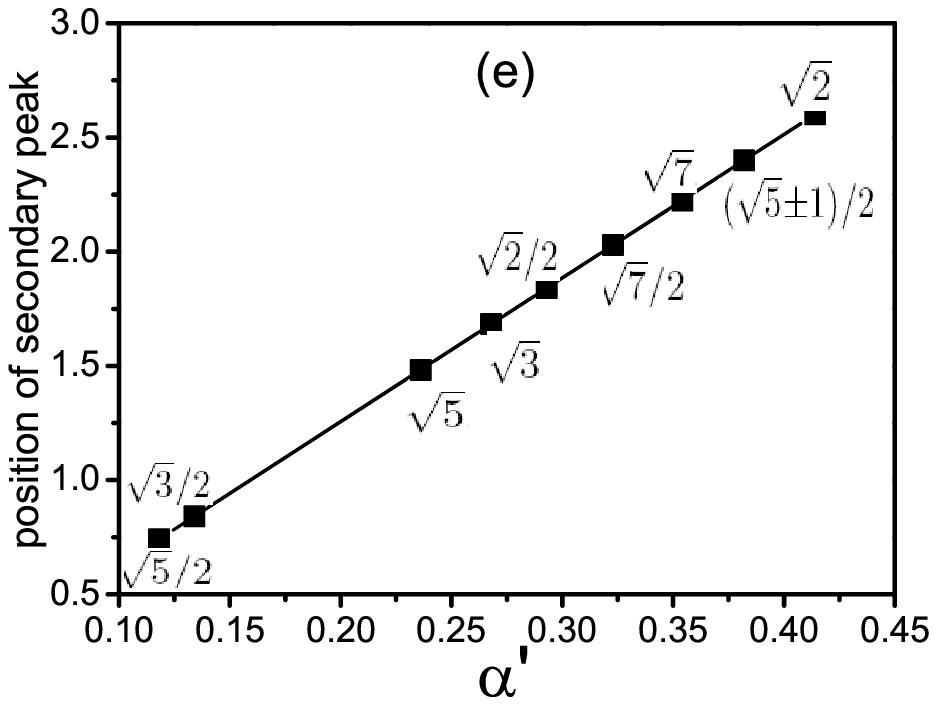}
\caption{Momentum distributions for systems with 1000 lattice
sites, $\alpha=(\sqrt{5}-1)/2$, 201 hard core bosons, and $V=1.1$
(a); $V=1.7$ (b); $V=2.3$ (c). (d): The value of secondary peaks
of momentum distributions relates to $V$ for systems with 1000
lattice sites, 201 bosons and $\alpha=(\sqrt{5}-1)/2$. (e): The
position of the secondary peak (the $k>0$ one) in the momentum
distributions relates to $\alpha'$ for systems with 1000 lattice
sites, 201 hard core bosons, and $V=1.1$. The irrational numbers
beside the points are the corresponding $\alpha$.} \label{Fig4}
\end{figure}

Then we consider the properties of the momentum distributions as the
strength of the incommensurate potential changing. In Fig.\ref{Fig4}
we show the momentum distributions for systems with three different
$V$. Low fillings are required for the existence of secondary peaks
in the momentum distributions. On increasing $V$, the value of
$n_{k=0}$ decreases which means that the coherence among the hard
core bosons decreases and the superfluid fraction decreases. Also
the peak at $k=0$ and the secondary peaks all become widespread. The
value of the secondary peaks (Fig.\ref{Fig4}d) first increases, and
it starts to decrease when $V$ is bigger than around $1.5$. Finally
it reaches an almost fixed number as the system going into the BG
phase. The population of high momenta states is always increasing
accompanying the decrease of the peak at $k=0$.

\begin{figure}[tbp]
\includegraphics[width=8.4cm, height=7.5cm, bb=25 20 303 235]{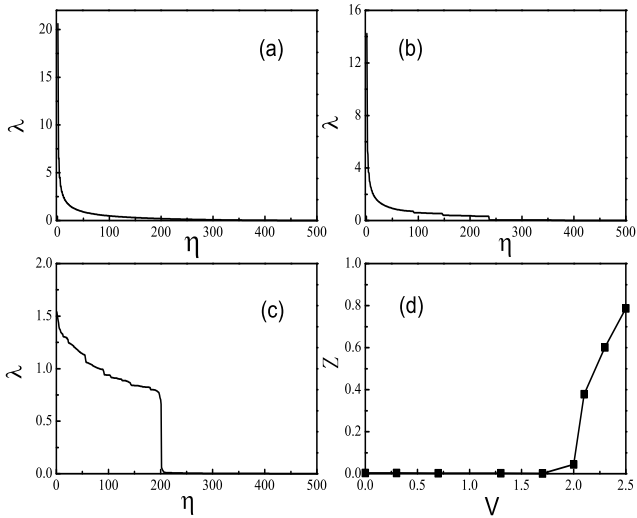}
\caption{Occupations of the natural orbitals for systems with 1000
lattice sites, 201 bosons, $\alpha=(\sqrt{5}-1)/2$ and $V=0$ (a);
$V=1.5$ (b); $V=2.3$ (c). (d): The amplitude of the
discontinuation ($Z$) defined as $Z=\lambda_N-\lambda_{N+1}$
relates to $V$ for systems with 1000 lattice sites, 201 bosons and
$\alpha=(\sqrt{5}-1)/2$.} \label{Fig5}
\end{figure}

Now we consider the properties of the secondary peaks. Their
positions are only decided by the value of $\alpha$ and are
irrelevant with the system size, $V$ and filling. But the
existence of the secondary peaks are related to the filling, and
the peaks only exist for low and high fillings. For the
incommensurate lattice with the form $\mathrm{cos}(2\pi\alpha i)$,
actually we can restrict $\alpha$ in the range of $[0,0.5]$
because $\mathrm{cos}[2\pi(\alpha-1)i]=\mathrm{cos}(2\pi\alpha i)$
and $\mathrm{cos}(-2\pi\alpha i)=\mathrm{cos}(2\pi\alpha i)$. For
example, $\alpha=(\sqrt{5}\pm1)/2$ are the same for the system. So
any $\alpha\in(-\infty,\infty)$ has an equivalent number in the
range $[0,0.5]$, and we denote it by $\alpha'$. The positions of
the secondary peaks are decided by the value of $\alpha'$. In
Fig.\ref{Fig4}e we show the position of the secondary peak (the
$k>0 $ one) as a function of $\alpha'$. From the data we can see
that the positions of the secondary peaks in the momentum
distribution are decided by $\alpha$ and at $k=\pm2\alpha'\pi$.
\begin{figure}[tbp]
\includegraphics[width=\linewidth, height=7cm, bb=25 20 303 235]{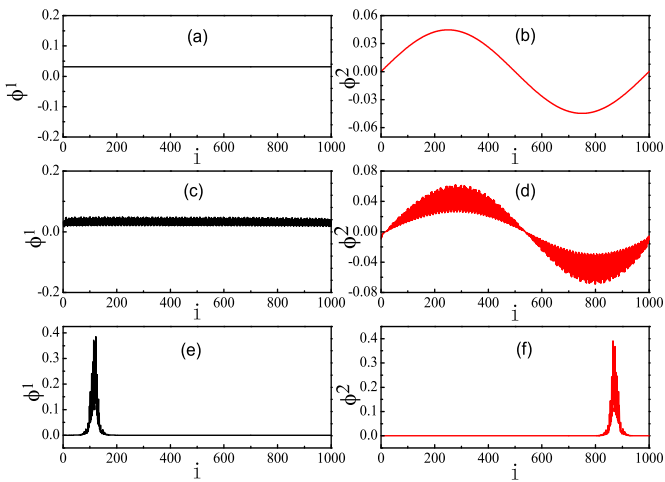}
\caption{Profiles of the two lowest natural orbitals for systems
with 1000 lattice sites, 201 bosons, $\alpha=(\sqrt{5}-1)/2$ and
$V=0$ (a)(b); $V=1.5$ (c)(d); $V=2.3$ (e)(f).} \label{Fig6}
\end{figure}

\begin{figure}[tbp]
\includegraphics[width=6.5cm, height=5cm, bb=25 20 303 235]{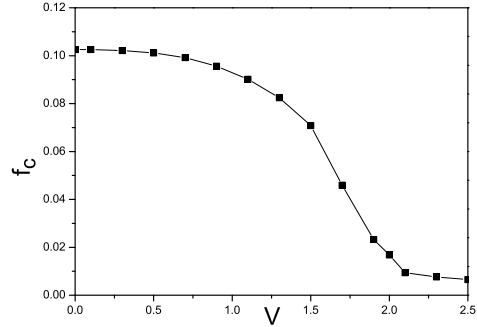}
\caption{The condensate fraction $f_c$ versus $V$ for systems with
1000 lattice sites, 201 bosons, and $\alpha=(\sqrt{5}-1)/2$.}
\label{Fig7}
\end{figure}

The natural orbitals $(\phi^\eta_i)$ are defined as the
eigenfunctions of the one particle density matrix \cite{Penrose},
\begin{equation}
\sum^L_{j=1}\rho_{ij}\phi^\eta_j=\lambda_\eta\phi^\eta_i,
\end{equation}
and can be understood as being effective one particle states with
occupations $\lambda_\eta$. For noninteracting bosons, all the
particles occupy in the lowest natural orbital and bosons are in the
BEC phase at zero temperature (only the quasi condensation exists
for the 1D hard core bosons). The occupations of the natural
orbitals for systems with three different $V$ are shown in
Fig.\ref{Fig5}. The occupations are plotted versus the orbital
numbers $\eta$, and ordered starting from the highest occupied one.
For $V<2$, the occupation distribution exhibits sharp single-peak
structure. The peak appearing in the lowest orbital is the feature
of the boson which is against the step function of the fermions.
With the increase in the strength of incommensurate potential, the
occupation of the lowest natural orbital ($\lambda_1$) decreases.
When $V>2$, no an obvious peak appears in the lowest natural
orbital. We also find that a discontinuation at $\eta=N$ emerges
when $V>2$. To characterize such a discontinuation, we define
$Z=\lambda_N - \lambda_{N+1}$ which indicates the occupation
difference between the $N$-th and $(N+1)$-th natural orbital for a
boson system with $N$ particles. The amplitude of the
discontinuation Z versus the strength of the incommensurate
potential ($V$) is plotted in Fig.\ref{Fig5}d. There is an obvious
change around $V=2$. For $V<2$ there is no discontinuation in the
occupation number. However, for $V>2$ a nonzero Z appears and the
amplitude of the discontinuation increases with the increase in $V$.

The effect of incommensurate potential on the natural orbital is
shown in Fig.\ref{Fig6}, where profiles of the two lowest natural
orbitals for the same system with three different $V$ are plotted.
When $V=0$, there is only the periodic optical lattice, and the
natural orbitals are plane waves. As $V$ increases but is smaller
than $2$, the natural orbitals still spread over all the lattice
corresponding to extended states, but there are a lot oscillations
in the waves induced by the existence of the incommensurate
potential which acts like random potential on sites because of the
irrational $\alpha$. When $V>2$ the states do not spread over all
the lattice any more and are localized. Correspondingly the system
is in the BG phase.

Finally we consider the influence of incommensurate potential on the
condensate fraction, which is defined as $f_c=\lambda_1/N$ to
indicate the ratio of occupation of lowest natural orbital. In
Fig.\ref{Fig7} we show the condensate fraction as function of $V$.
For small $V$ the condensate fraction $f_c$ decreases slowly with
the increase in $V$. As $V$ increases further to approach $V=2$, the
condensate fraction decreases rapidly. For $V>2$ there is almost no
condensation. The change of the condensate fraction also gives
signature of the superfluid to insulator transition in the
incommensurate optical lattice system.



\section{Summary}
In summary, we have studied the properties of hard core bosons in an
incommensurate optical lattice. Using the Bose-Fermi mapping and the
exact numerical method proposed by Rigol and Muramatsu \cite{Rigol},
we exploit the phase transition from superfluid to the localized BG
phase as the strength of the incommensurate potential increases from
weak to strong. We calculate the superfluid fraction, one particle
density matrices, momentum distributions, the natural orbitals and
their occupations. All of these quantities show that there exists a
phase transition in the system when the strength of incommensurate
potential exceeds $V=2$. Our study provides an exact example which
unambiguously exhibits the transition from superfluid to Anderson
insulator in an incommensurate optical lattice.

\begin{acknowledgments}
This work was supported by NSF of China under Grants No.10821403 and
No.10974234, programs of Chinese Academy of Science, 973 grant
No.2010CB922904 and National Program for Basic Research of MOST.
\end{acknowledgments}

\end{document}